# Coupled Negative magnetocapacitance and magnetic susceptibility in a Kagomé staircase-like compound $Co_3V_2O_8$.

Natalia Bellido, Christine Martin, Charles Simon, Antoine Maignan

Laboratoire CRISMAT, UMR 6508, ENSICAEN-CNRS,

6 Boulevard du Maréchal Juin, 14050 Caen Cedex, France

The dielectric constant of the Kagomé staircase-like $Co_3V_2O_8$ polycrystalline compound has been measured as function of temperature and magnetic field up to 14T. It is found that the application of an external magnetic field suppresses the anomaly for the dielectric constant beyond 6.1K. Furthermore, its magnetic field dependence reveals a negative magnetocapacitance which is proportional to the magnetic susceptibility, suggesting a common magnetostrictive origin for the magnetic field dependence of the two quantities. This result is very different from that obtained from the isostructural compound $Ni_3V_2O_8$ that presents a peak in the dielectric constant at the incommensurate magnetic phase transition coupled to a sign change of the magnetocapacitance.

## I. Introduction

Developing new multifunctional materials for electronics of the future is a challenging goal for solid-state chemists and physicists[1]. This kind of research has been illustrated by the multiferroic materials and in particular those with a coupling between ferroelectricity and ferromagnetism[2]. Such materials are developed for multiple-state-memory elements based on writing of ferroelectric data and reading the induced magnetic field change[3]. Another class of multifunctionality is shown by the materials with magnetocapacitance effects[4], i.e. which dielectric constant ($\varepsilon$) depends on application of an external magnetic field. In the search of magneto electric effect during the 60's, Schmid proposed a classification of systems in terms of symmetry[5]. Since ferromagnetic order does not allow time-inversion symmetry such as spatial polarization inversion symmetry, the number of systems showing both ferromagnetic and spatial polarization is highly restricted[6,1]. However, the recent discovery of new multiferroic materials, in which ferroelectricity is induced by magnetic order, has revived the research into magneto electric coupling. These systems are frustrated spin-density-waves magnets in which magnetic order breaks inversion symmetry leading to ferroelectricity when spin rotation axis is not parallel to the wave vector[7], as in $RMnO_3$ (R= Gd, Tb, Dy)[8], $RMn_2O_5$ (R=Tb, Ho, Er, Dy)[9,10,11,12] or $Ni_3V_2O_8$[13].

The compound $Ni_3V_2O_8$[13] is an extensively studied system belonging to this class of multiferroicity. Magnetism in this compound comes just from the S=1 spin of $Ni^{2+}$ cations that are distributed in a staircase-like Kagomé lattice[14]. This structure presents two inequivalent sites for $Ni^{2+}$ and magnetic frustration, leading to complex (H,T) magnetic phase diagram[15]. Four magnetic phase transitions were revealed by specific heat, magnetic measurements[16] and neutron diffraction[17]. The first kind of magnetic order appears below 9K consisting of a high temperature incommensurate magnetic structure (HTI) described by modulated amplitude in which only one of the two nickel sites is concerned. Then, cooling down further, at T=6.3K, another low temperature incommensurate structure is found (LTI), described as a helical order concerning both nickel sites. For the latter, the breaking of inversion symmetry induces the charge displacement responsible for a polarization along the b-axis[18] that can be controlled via an external magnetic field[13]. Below the temperature of appearance for these incommensurate phases, two canted antiferromagnetic commensurate phases take place with a weak ferromagnetic moment along the crystallographic axis c at T=4K (C') and T=2.3K (C).

Two other isostructural magnetic compounds have been synthesized $Co_3V_2O_8$[19] and $\beta$-$Cu_3V_2O_8$[20] with spin S= $3/2$ and S= $1/2$, respectively. In this paper we report on the study of the magnetic and dielectric properties of the spin S= $3/2$ compound, $Co_3V_2O_8$, also characterized by four magnetic phase transitions[21]. These results are compared to the $Ni_3V_2O_8$ ones.

## II. Experimental

The polycrystalline sample of $Co_3V_2O_8$ has been prepared by solid-state reaction in a silica tube. The $Co_3O_4$, $V_2O_5$ and $V_2O_3$ in the 1:0.5:0.5 ratios were mixed in an agate mortar. After pressing, the sample was put in finger-likes alumina crucible which was set in ampoule. The latter, sealed under primary vacuum was heated at 1100°C for 12 hours and then quenched in air. The structural characterization of the obtained black bars of ceramics was made by X-ray powder diffraction by an X-pert Pro diffractometer. All the diffraction lines could be indexed in the orthorhombic structure of $Co_3V_2O_8$ (space group Cmca) in good agreement with ref[16]. The corresponding structure with the Kagomé staircase-like structure is given in figure 1.

To obtain the values for the dielectric constant, complex impedance measurements were performed using a commercial AG4284A LCR-meter. In order to measure in a magnetic field, a complete sample holder with four coaxial cables was designed. This allows measurements of the dielectric constant to be made inside a Quantum Design PPMS. The magnetization was measured with the magnetic option (extraction method) of the PPMS. The electrical contacts were deposited by an ultrasonic method with indium solder covering two parallel faces of the sample in order to approximate the geometry of a parallel plate capacitor. For accuracy an ac-voltage of 1V at a frequency of 100 kHz was applied to the sample. Nonetheless, a systematic check was also performed to rule out any frequency and amplitude dependence of the results. In particular, the frequency dependence of the complex impedance allows modeling our samples as a capacitor and a resistance in parallel. The dielectric constant has been extracted from the value of the capacitance by using the sample dimensions.

## III. Results and discussion

Recent magnetization and heat capacity studies[21] of a single crystal of $Co_3V_2O_8$ have shown four magnetic phase transitions at 11.2K, 8.8K, 6.6K and 6.1K. Following the magnetic phase diagram of the isostructural compound $Ni_3V_2O_8$, the same progression of magnetic phases was proposed[21]. This interpretation suggests an equivalent ferroelectric LTI phase between 8.8K and 6.6K, and the presence of transitions with a peak in the dielectric constant. There was no polarisation measurement in $Co_3V_2O_8$ and we did not measure this property neither. Neutron scattering study shows that incommensurate phases occur below 6.6K (LTI) and between 8.8K and 11.2 K (HTI)[22]. In the same paper, preliminary measurements of the dielectric constant as function of temperature show no peak at 8.8K, but rather an abrupt change at ~ 6K. We have reproduced this anomaly (figure 2 upper panel). The application of magnetic field softens the anomaly that extends its temperature range of observation up to high temperatures, around 30K (fig.2, bottom panel), while no effect is observed below 6K. Furthermore this anomaly is completely suppressed by application of a sufficiently large magnetic field (≥4T). The behaviour is reminiscent of the anomaly observed in the ferroelectric hexagonal $RMnO_3$ (R= Y, Lu, Sc)[23,24] when entering in the antiferromagnetic phase, and in the ferroelectric perovskite compound $BiMnO_3$ when crossing the ferromagnetic transition[4].

Isothermal curves of the magnetic field dependence of the dielectric constant have been also collected. Below 6 K, no effect has been observed (not shown). This was interpreted as a ferromagnetic phase where no effect of the magnetic field is observed[22]. In contrast, for T>6K, very interesting data have been obtained. We observed three types of behaviour presented in figure 3: For temperatures above 30K (illustrated by T=50K), a quadratic dependence is observed, similar to what was observed in $YMnO_3$[25] and $BiMnO_3$[4]. As the temperature is decreased, the shape evolves continuously towards a peak centred at H=0, which width decreases (illustrated by 20K). Below 8.8K, the peak becomes very sharp (illustrated by 7K). Remarkably, whatever the temperature is, this evolution follows the magnetic field dependence of the magnetic susceptibility ($\chi = dM/dH$) that is also shown in the same figure (right y-axis).

The similar trend observed for the dielectric constant and the magnetic susceptibility suggests a common origin. According to the strong coupling between spins, charges and the lattice often found in transition metal oxides, it is proposed that the application of the magnetic field induces a continuous structural distortion, responsible for the changes in both magnetization and dielectric susceptibility.

Let us start the discussion of the magnetic susceptibility by the measurement of its temperature dependence (figure 4). The data are very well fitted by a Curie law giving θ=4.4K and 5.14$\mu_B$ in agreement with the previous data[21]. For 50K and 20K, the magnetic field dependence of the susceptibility in the paramagnetic state follows a Langevin's law where the important control parameter is: $g\mu_B H / k_B(T-T_c)$, which exhibits a dome shape as function of magnetic field. At 50K, only the top of the dome is visible, and, at 20K, the entire bell shape is observed. In contrast, at lower T (below 8.8K) there is sharp peak at H=0 associated with a sharp transition from the zero field frustrated antiferromagnetic phase[22] to a ferromagnetic component reported by magnetization[21]. In the paramagnetic phase, it is particularly clear that the Zeeman energy and the spin-orbit coupling on a single magnetic site are the only terms which are relevant and are at the origin of the magnetic field variation of both magnetic and dielectric susceptibilities.

This is very different from the dielectric susceptibility of $Ni_3V_2O_8$ which is shown in figure 5 in the incommensurate phases above 4K. The peak which is observed at H=0 corresponds to the LTI-HTI transition[20]. Under the application of an external magnetic field, the peak broadens and shifts to lower temperatures following the magnetic transition as already reported[15] up to 8T. Above 8T, the new data reveal that the peak broadens very rapidly and disappears. In order to follow the corresponding phase transition, the evolution of the dielectric constant as a function of the magnetic field is recorded (fig. 6). We have found a positive magnetocapacitance below $T_{LTI}$ (6.3K) while a negative one is found above. Above 7K (figure 5), the magnetocapacitance is again positive, decreasing slowly up to 20K in the paramagnetic phase. At 4.5K, 5K and 6K, the LTI-HTI transition appears at about 8.5T as a cusp (figure 6). Small additional features at low field (1-2T) appear at 4.5K and 5K which do not correspond to any reported magnetic transition.

### III. Conclusions

The present result show that the magnetocapacitance of the two isostructural compounds ($Co_3V_2O_8$ and $Ni_3V_2O_8$) are very different. In the Ni-containing compound, the majority of the magnetic phases is paraelectric and the only LTI phase is ferroelectric. The dielectric constant exhibits a peak at the LTI-HTI magnetic phase transition. The magnetocapacitance is positive in the LTI and HTI phases with a sign change observed in the region of the LTI-HTI transition. In contrast, the compound $Co_3V_2O_8$ shows always a negative magneto-capacitance which is proportional to the magnetic susceptibility. This suggests strongly that a structural distortion is induced by the magnetic field. The existence of a strong spin-orbit coupling in the case of octahedral coordination of the high spin $Co^{2+}$ could play a crucial role.

Acknowledgements: Natalia Bellido acknowledges financial support to NOVELOX ESRT MEST-CT-2004-514237 and CNRS.

**Figure 1: The structure of $Co_3V_2O_8$, showing the Kagomé staircase-like structure. In blue and green, the two types of Co atoms. Oxygen and vanadium atoms are not shown.**

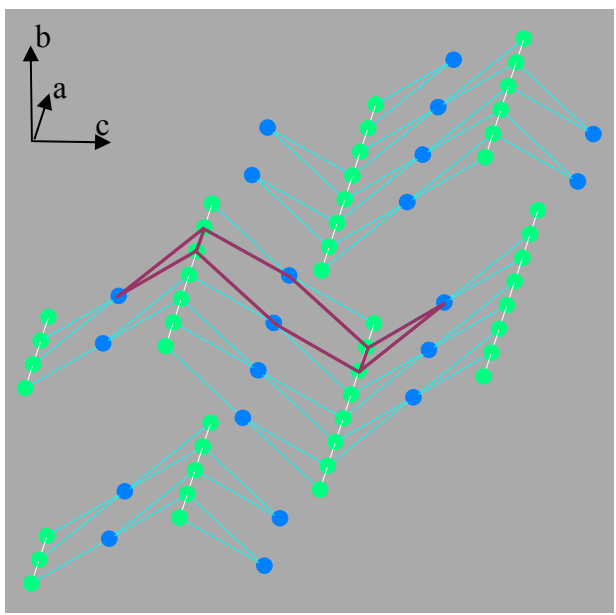

**Figure 2.** Temperature dependence of dielectric constant under different magnetic fields for Co$_3$V$_2$O$_8$. Top part, an enlargement of the low temperature part. Bottom part, a general view of the magnetic field dependence.

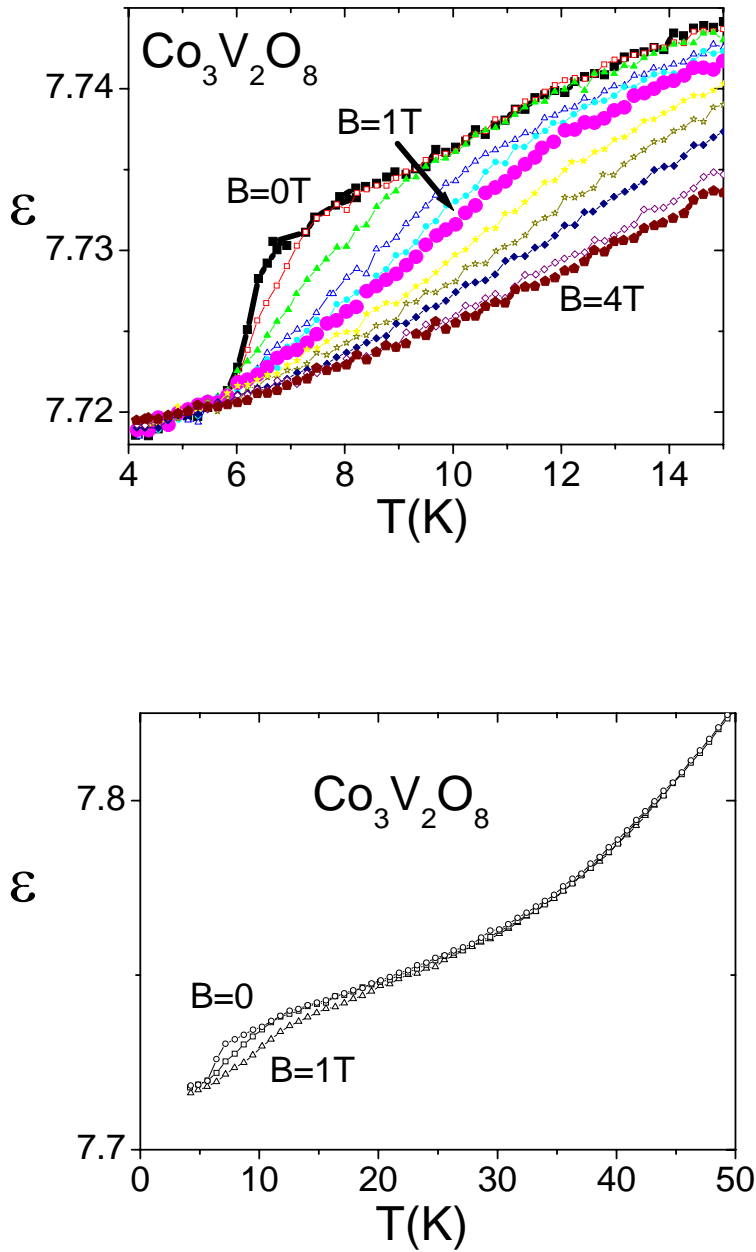

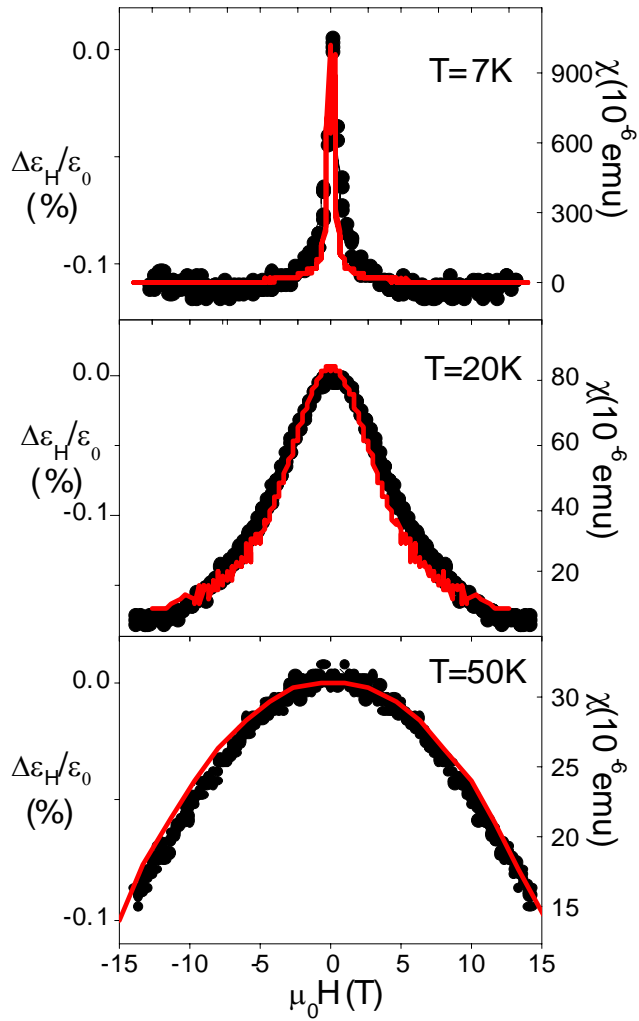

**Figure 3. Magnetic field dependence of the relative variation of the dielectric constant at different temperatures for $Co_3V_2O_8$. Left axis (black dots): $\Delta\varepsilon/\varepsilon_0$ is defined by $(\varepsilon(H)-\varepsilon(0))/\varepsilon(0)$. Right axis (red lines): the magnetic susceptibility is also reported for comparison.**

**Figure 4:** Temperature dependence of the magnetic susceptibility registered at 10Oe (1000Hz) of $Co_3V_2O_8$. A fit by a Curie law (in red) is also presented with the two parameter values θ=4.4K and $\mu_{eff}$=5.14 $\mu_B$ for the high temperature part of the curve (above 15K).

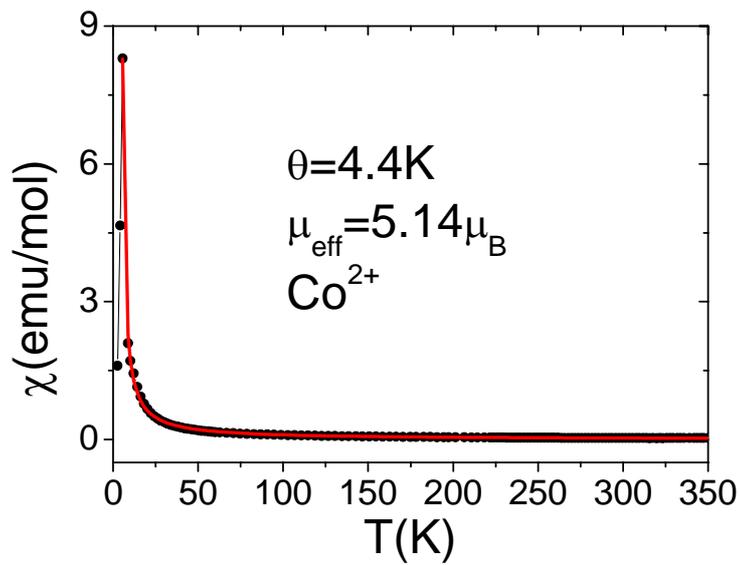

**Figure 5.**

**Temperature dependence of dielectric constant under different magnetic fields for Ni$_3$V$_2$O$_8$. The peak is attributed to the ferroelectric transition at T$_{LTI-HTI}$.**

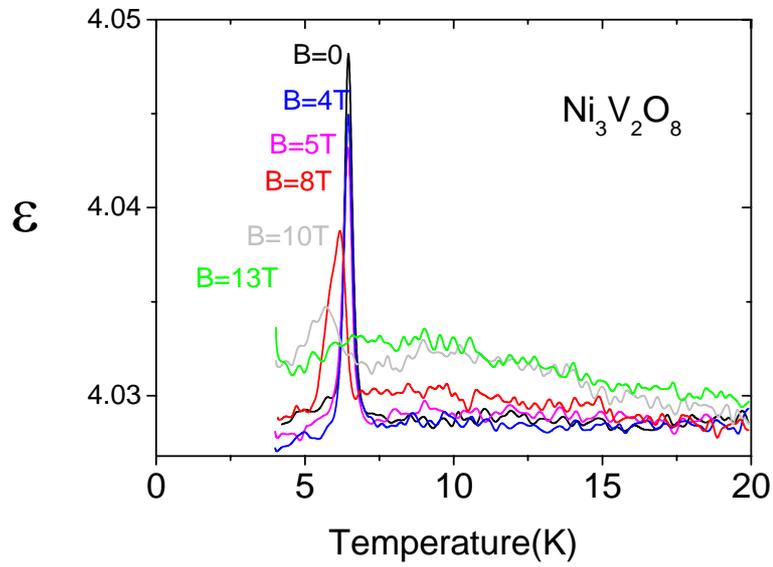

**Figure 6.**

Magnetocapacitance of $Ni_3V_2O_8$ for different temperatures, showing positive magnetocapacitance in the LTI and HTI phases and negative in the region of the phase transition.

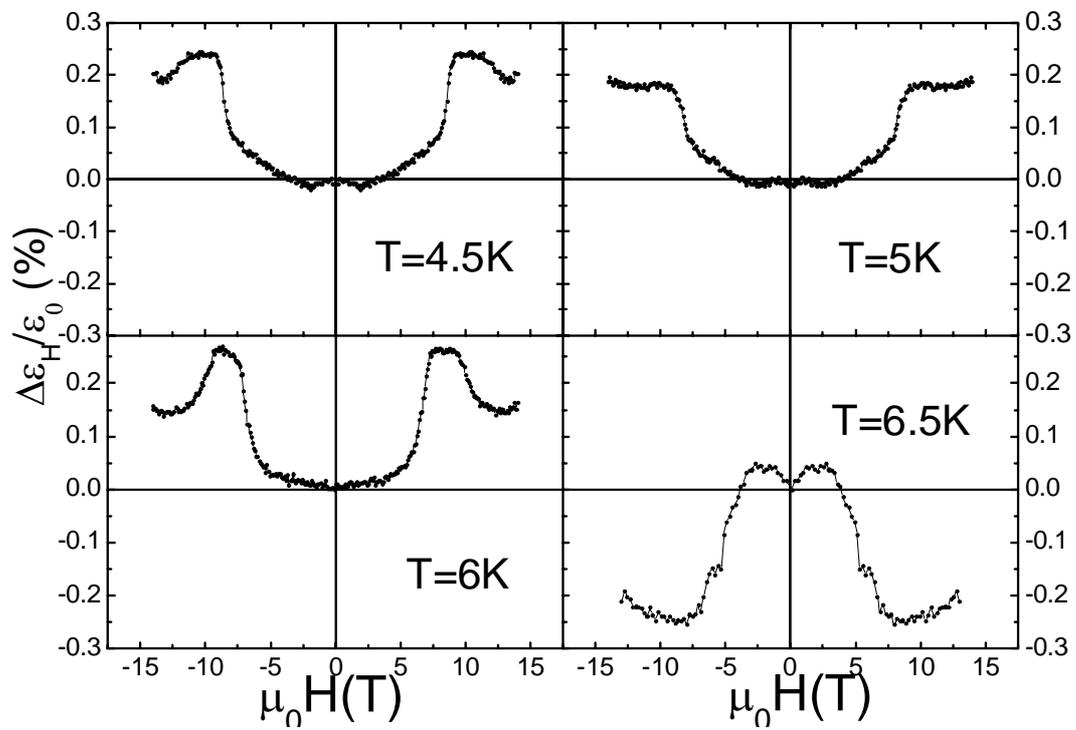